\documentclass[10pt,twocolumn,letterpaper]{article}

\usepackage[pagenumbers]{wacv} 

\usepackage{graphicx}
\usepackage{amsmath}
\usepackage{amssymb}
\usepackage{booktabs}
\usepackage{tabularx}

\usepackage[pagebackref,breaklinks,colorlinks]{hyperref}

\usepackage[capitalize]{cleveref}
\crefname{section}{Sec.}{Secs.}
\Crefname{section}{Section}{Sections}
\Crefname{table}{Table}{Tables}
\crefname{table}{Tab.}{Tabs.}

\def\wacvPaperID{05} 
\def\confName{WACV}
\def\confYear{2025}

\begin{document}

\title{“What's Happening”- A Human-centered Multimodal Interpreter Explaining the Actions of Autonomous Vehicles}

\author{
Xuewen Luo$^{1,\dagger}$, Fan Ding$^{1,\dagger}$, Ruiqi Chen$^{2,\ddagger}$, Rishikesh Panda$^{3,\ddagger}$, Junnyong Loo$^{1}$, Shuyun Zhang$^{4,*}$ \\
$^1$School of Information Technology, Monash University Malaysia, Bandar Sunway, Selangor, Malaysia \\
$^2$Paul G. Allen School of Computer Science and Engineering, University of Washington, Seattle, WA, USA \\
$^3$BITS Pilani K K Birla Goa Campus, Goa, India \\
$^4$Department of Digital Humanities, King's College London, London, United Kingdom
}

\maketitle

\begin{abstract}
Public distrust of self-driving cars is growing. Studies emphasize the need for interpreting the behavior of these vehicles to passengers to promote trust in autonomous systems. Interpreters can enhance trust by improving transparency and reducing perceived risk. However, current solutions often lack a human-centric approach to integrating multimodal interpretations. This paper introduces a novel Human-centered Multimodal Interpreter (HMI) system that leverages human preferences to provide visual, textual, and auditory feedback. The system combines a visual interface with Bird’s Eye View (BEV), map, and text display, along with voice interaction using a fine-tuned large language model (LLM). Our user study, involving diverse participants, demonstrated that the HMI system significantly boosts passenger trust in AVs, increasing average trust levels by over 8\%, with trust in ordinary environments rising by up to 30\%. These results underscore the potential of the HMI system to improve the acceptance and reliability of autonomous vehicles by providing clear, real-time, and context-sensitive explanations of vehicle actions.
\end{abstract}

\makeatletter
\renewcommand{\@makefntext}[1]{\noindent\makebox[1.8em][r]{\@thefnmark.}#1}
\makeatother

\footnotetext[1]{$^{\dagger}$These authors contributed equally as first authors.}
\footnotetext[2]{$^{\ddagger}$These authors contributed equally as second authors.}
\footnotetext[3]{$^{*}$Corresponding authors.}

\section{Introduction}
In recent years, public distrust of autonomous vehicles (AVs) has significantly increased. A 2023 survey by AAA highlighted that fear of self-driving cars is on the rise \cite{aaa_2023}. To address this growing concern, current research has proposed various strategies to enhance public trust and acceptance of AVs, such as improving user experience \cite{walker2023trust}, addressing ethical considerations \cite{chen2023research}, and conducting comprehensive safety testing \cite{katiyar2024ai}. 

One promising approach to improving user trust is the use of a driving behavior interpreter, which has been identified as effective in enhancing system transparency, reducing perceived risk, and increasing acceptance \cite{choi2015}. Research indicates that explanations can boost trust and mitigate negative emotional responses to unexpected vehicle behaviors \cite{omeiza2021}\cite{wiegand2020}. Our preliminary study also supports this, with 90\% of testers indicating that their trust in autonomous driving(AD) would increase if there were a system to explain driving behaviors.

Currently, behavioral interpretation methods encompass various forms, including textual, visual, and speech interpretations \cite{phongphaew2024text}\cite{ruijten2018enhancing}\cite{du2018voice}. However, most existing approaches rely on a single modality for explaining driving behavior. Our preliminary study reveals that 56.67\% of users prefer having the option to choose their preferred interpretation method. This finding underscores a significant demand for an integrated multimodal interpreter. Furthermore, existing interpreters predominantly concentrate on providing technical explanations \cite{ribeiro2016}\cite{choi2015}\cite{dong2023}, often neglecting the human-centric factors essential for improving trust and acceptance among users.

Therefore, there is an urgent need for a human-centered, multimodal automated driving behavior interpreter that not only integrates multiple modalities but also provides user-friendly explanations. To address this need, we propose a Human-centered Multimodal Interpreter (HMI) system that combines various presentation modalities and employs prompt engineering to fine-tune a Large Language Model (LLM). The HMI system integrates a visual interface featuring Bird’s Eye View (BEV), maps, and text displays with voice interaction facilitated by the fine-tuned LLM. By leveraging the LLM's ability to generate human-like explanations, the system provides real-time, concise, and comprehensible responses to user queries. This multimodal integration, along with the LLM's human-centric explanations, aims to enhance users' trust and acceptance of autonomous vehicle technology by offering timely, diverse, and easily understandable explanations of vehicle actions.

User study examines the HMI system's effectiveness in improving passenger trust and its varying impacts across different scenarios. The implementation of the HMI system significantly increased user trust in AVs under three driving conditions (Ordinary Environment, Low Visibility, and Emergency Braking), with average trust levels rising by over 8\%. Notably, our system increased user trust in ordinary environments by over 30\%.

This study makes the following contributions:

\begin{itemize} 

\item \textbf{Novel HMI for Autonomous Driving}: A HMI system is designed to provide explanations to passengers in autonomous driving contexts. This system integrates visual, auditory, and textual feedback to enhance passenger understanding and trust.

\item \textbf{Enhanced LLM Prompt Design}: An advanced prompt system is developed to improve the text output of a LLM. This enables the LLM to provide context-specific and precise explanations, tailoring its responses based on different driving scenarios.

\item \textbf{Empirical Validation of HMI Effectiveness}: The effectiveness of the HMI interpreter is empirically validated through experiments. The results demonstrate that the HMI system significantly improves passenger trust, with the most pronounced effects observed in ordinary environments. 

\end{itemize}

\section{Related Work}
\subsection{The Needs for Explanations in Autonomous Driving}

Public distrust of AVs has risen significantly in recent years. According to a 2023 survey by AAA, establishing a mechanism for trust in autonomous vehicles is essential. \cite{aaa_2023}. To enhance public trust in AVs, current research has proposed various strategies focusing on user experience\cite{walker2023trust}, ethical considerations\cite{chen2023research}, and comprehensive safety testing\cite{katiyar2024ai}\cite{ding2024energy}. These approaches are crucial for addressing public concerns and increasing the acceptance of AVs technology.

Providing clear explanations of AVs actions and decisions reduces the mystery of AVs operations, making the technology more understandable, approachable, and trustworthy\cite{atakishiyev2021explainable}. Recent research supports the idea that equipping AVs with driving behavior interpreters can improve user trust by enhancing system transparency and reducing perceived risk \cite{omeiza2021}. Studies demonstrate that explanations significantly increase trust and mitigate negative user reactions, highlighting the crucial role of clear and informative explanations for user acceptance of AVs \cite{taylor2023reliable}.

To enhance trust in automated driving systems, it is essential to provide explanations that are clear, comprehensible, and aligned with user preferences. Effective methods for providing explanations include simple and easily understandable textual descriptions\cite{phongphaew2024text}, intuitive visual interface\cite{ruijten2018enhancing}, and convenient voice explanations\cite{du2018voice}. Furthermore, multimodal feedback plays a key role in improving user experience (UX). Gomaa et al. (2022) found that users often prefer a combination of feedback modalities, suggesting a preference for multisensory information \cite{gomaa2022adaptive}. Integrating visual, auditory, and haptic elements into explanations has the potential to cater to diverse user needs and learning styles, ultimately enhancing trust and acceptance of AVs \cite{nakagawa2016multimodal}\cite{sun2021improvement}\cite{luo2024pkrd}. 

Systems that interpret driving behavior in AVs enhance transparency and reliability, fostering user trust. Current methods often use single-modal approaches, missing comprehensive explanations. Our study integrates multiple modalities for a more user-friendly system, improving AVs transparency, reliability, and user acceptance.

\subsection{Existing Interpreters in Autonomous Driving}

Current AVs interpreter models predominantly provide technical explanations tailored for engineers, rather than human-centric explanations for passengers \cite{atakishiyev2024explainable}. For example, Ribeiro et al. (2016) introduced the Local Interpretable Model-agnostic Explanations (LIME) model \cite{ribeiro2016}, which effectively explains classifier predictions. However, while LIME is valuable for technical debugging and understanding, it lacks the user-friendly interpretations that everyday users need to comprehend AVs decisions intuitively.

Similarly, Dong et al. (2023) highlight the critical need for explainable AI in AVs but maintain a strong focus on the technical aspects of implementation \cite{dong2023}. Their work underscores the importance of transparency and interpretability to foster trust and acceptance of AVs technology among users. Nevertheless, their approach primarily caters to a technical audience, overlooking the necessity for explanations that are easily understandable to non-technical users.

Recognizing the limitations of purely technical explanations, studies have highlighted the importance of human-centered explanations. Choi and Ji (2015) argue that building trust in AVs is important for their widespread adoption. They contend that technical explanations alone are insufficient for this purpose \cite{choi2015}. Instead, they advocate for explanations that are intuitive and easily understandable to non-expert users, as these can significantly enhance user trust and acceptance \cite{atakishiyev2024explainable}. 

Current explanatory models for autonomous driving primarily offer technical details for engineers, but there is growing recognition of the need for human-centered explanations to build trust and acceptance among everyday users. Research is increasingly shifting towards interactive and multimodal approaches to improve transparency and user-friendliness. Our research aims to address this by prioritizing human-centered explanations and leveraging multimodal interaction to bridge the gap between technical explanations and user needs in AVs.

\section{Preliminary Study}
To guide the design of a human-centered multimodal interpreter  system, it is essential to first understand current user expectations. Previous research has provided some insights into passenger attitudes and specific concerns regarding autonomous vehicles.Suggesting that interpreters are beneficial, many studies fail to specify the types of interpreters, their methods of assistance, and the contexts in which they are needed, leading to an inability to accurately capture customer needs.

To address this question, we conducted an online survey with 30 English-speaking participants who possessed varying levels of knowledge or awareness of autonomous driving. The gender distribution of our survey participants was approximately equal, with 58\% identifying as male and 42\% as female. The age distribution was relatively balanced, with an average age of 33.8 years. All participants are regular drivers or passengers in their daily lives and were selected based on their prior exposure to autonomous driving technology. This includes direct experiences (e.g., riding in autonomous vehicles) and indirect experiences (e.g., extensive learning or reading about the technology). This criterion ensured that the participants had a fundamental understanding of autonomous driving systems, allowing us to gain more expert insights into the perception of autonomous driving.

The survey was designed to provide a comprehensive understanding of users' perceptions, concerns, and expectations regarding autonomous driving. Participants responded to a series of questions, including multiple-choice, Likert scale, and open-ended formats. For example, questions covered topics such as: Interactive Influences—“To what extent does interacting with the AVs help you feel safer and trust the vehicle?”; Explanation Demands—“If the AVs explains why it is steering or decelerating, will this increase your trust in the vehicle?”; and Best Modes—“Which modality (voice explanation, visual feedback, textual explanation, or a combination of these) do you think would best enhance your trust in the AVs?” 

The following are the features of the autonomous driving interpreter that users most desire, as summarized by the survey:

\textbf{1. Transparency and Interpretability}:

Users strongly prefer systems that offer clear and understandable explanations for decisions and actions. The survey revealed that 21 out of 30 participants (70\%) felt uneasy when they could not understand an autonomous vehicle's behavior, particularly during unexpected actions. Moreover, 27 participants (90\%) believed that their trust in the autonomous vehicle would improve significantly if it could explain its actions. Participants expressed that their trust would increase if the AVs could explain its actions, such as the reasons for steering or slowing down. 
For instance, P\textsubscript{12} highlighted the importance of understanding specific actions, stating\textit{"the car turned left to avoid colliding with a parked car" helped participants understand and trust the AV’s decisions.} P\textsubscript{15} reinforced this sentiment by emphasizing \textit{"transparency and explainability are crucial for building trust."} Additionally, other participants, such as P\textsubscript{7} and P\textsubscript{22}, noted that

\textit{"Providing precise explanations for the AV's behavior allows users to grasp the system’s operational logic, thereby increasing their trust."} by P\textsubscript{7}.

\textit{"Such explanations not only alleviate confusion and anxiety but also enhance users' sense of control and security over the AV."} by P\textsubscript{22}.

\textbf{2. Real-time Feedback}:

Participants expect the system to provide immediate and relevant feedback during operation. The survey results indicate that most participants believe interacting with the AVs significantly enhances their sense of security and trust, with ratings predominantly between 4 and 5.

Real-time feedback emerged as a crucial factor in improving user understanding and predicting system behavior. This is emphasized by P\textsubscript{9} as \textit{"Real-time feedback is crucial for improving user understanding and predicting system behavior,"} Similarly, P\textsubscript{14} mentioned \textit{"When users have real-time access to information about the AV’s status and actions, they can better grasp the current driving environment and the basis for the system’s decisions."} 

\textit{"This feedback mechanism not only helps users adjust their expectations promptly but also boosts their confidence and reliance on the AVs,"} explained by P\textsubscript{22}.

These insights indicate that when users have real-time access to information about the AV’s status and actions, they can better understand the current driving environment and the rationale behind the system’s decisions. This feedback mechanism not only helps users adjust their expectations promptly but also boosts their confidence and reliance on the AVs.

\textbf{3.Personalization}:

The system's ability to adapt its behavior and responses to individual preferences and needs was a highly valued feature according to the survey. There are 17 of 30 participants agreed that personalized settings enhance their trust and satisfaction with AVs. They mentioned that they would be more inclined to accept and use an AVs if the system could be customized to their specific needs and preferences for feedback. For example, P\textsubscript{3} stated \textit{"The ability to choose how I receive feedback from the AVs would make me feel more comfortable and in control."} Similarly, P\textsubscript{7} mentioned\textit{"As someone with a hearing impairment, I would trust the AVs more if it provided visual alerts."} P\textsubscript{15} noted that \textit{"Personalization features that cater to my specific needs, like providing audio explanations, would definitely increase my trust in the system."} P\textsubscript{20} explained that \textit{"Allowing users to select how feedback is provided based on their preferences makes the system more relevant and user-friendly."}

These insights indicate that personalization features make the system more relevant and user-friendly. By allowing users to select how feedback is provided based on their preferences, AVs can offer a more tailored interaction experience. This customization not only boosts user satisfaction but also strengthens their trust and reliance on the system.

\section{Our HMI System}
Users interact with the HMI system through both visual screens and voice communication. The system processes sensory data from autonomous vehicle, executes interpretation tasks and presenting the information through bird's-eye view (BEV), maps, text, and voice outputs. This ensures that information is effectively communicated and easily accessible. The user experience involves automated interpretation, user-initiated queries or commands, and the receipt of multimodal responses. Additionally, users can choose how to get feedback to further optimize the interaction and enhance overall trust in the autopilot system.

\subsection{Working Mechanism of HMI System}
The HMI system is an innovative solution designed to meet the user's need for explanations in the scenario of riding in a self-driving car. We understand that users expect easy to understand, real-time and preferred explanations to increase their confidence and reduce discomfort.

\begin{figure}[ht]
    \centering
    \includegraphics[width=0.7\columnwidth]{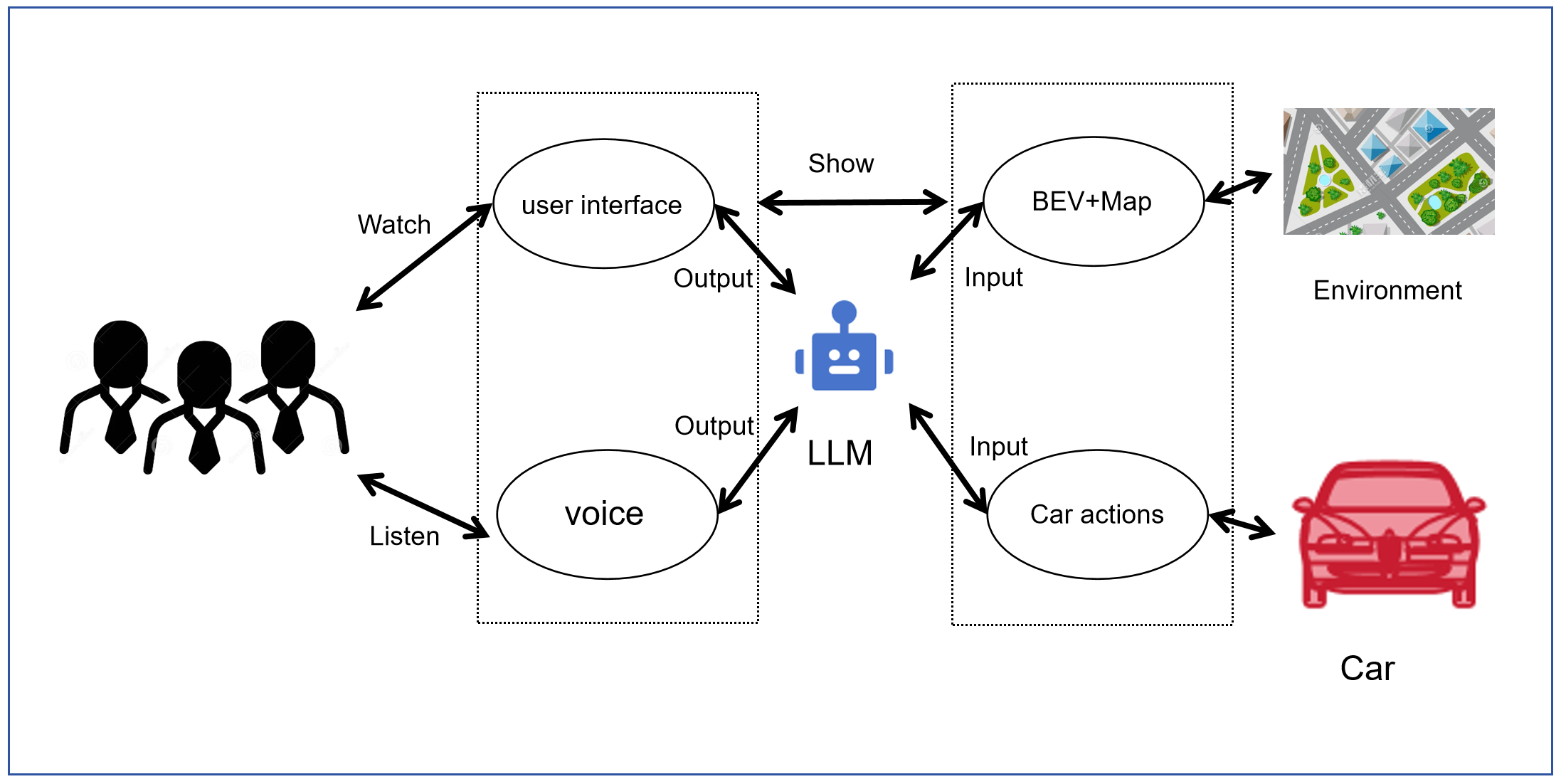}
    \caption{Working Mechanism of HMI System}
    \label{visual interface}
\end{figure}

The HMI system integrates multimodal components, including Bird's Eye View (BEV), maps, and text displays, to provide real-time visual feedback. This setup is implemented using AirSim \cite{shah2018airsim}, which provides a high-fidelity driving environment for accurate environmental visualization. Additionally, the system employs advanced voice interaction technology powered by a Large Language Model. We used the GPT-4 model as the LLM and prompt engineering fine-tuning \cite{xu2023making} to ensure the clearest and most concise language in its speech output. This technology leverages the LLM’s human-like language capabilities to explain complex autonomous driving behaviors, enabling passengers to comprehend the vehicle’s specific actions and decision-making processes. To minimize auditory noise for the passenger, the LLM has been fine-tuned to deliver only the clearest and most concise language in its speech output.

To meet the diverse needs of passengers and improve system adaptability, we have provided options for both visual (images and text) and auditory feedback. This multimodal interpreter ensures that passengers can choose their preferred method of receiving information, especially accommodating the specific needs of individuals with disabilities.

\subsection{Visual Interface with BEV, Map, and Text Display}
In designing the user-friendly visualization interface for the HMI system, we prioritized user expectations. Our core design focus was on providing clear and easily understandable information. We ensured that the interface is simple and intuitive, allowing users to quickly access the information they need.

In our HMI systems, the visual interface \textbf{Figure \ref{visual interface}} is designed to provide a full and intuitive representation of the vehicle's driving status and surroundings. The interface emphasizes three key modules: Bird's Eye View, Map, and Text Display.

\begin{figure}[ht]
    \centering
    \includegraphics[width=0.7\columnwidth]{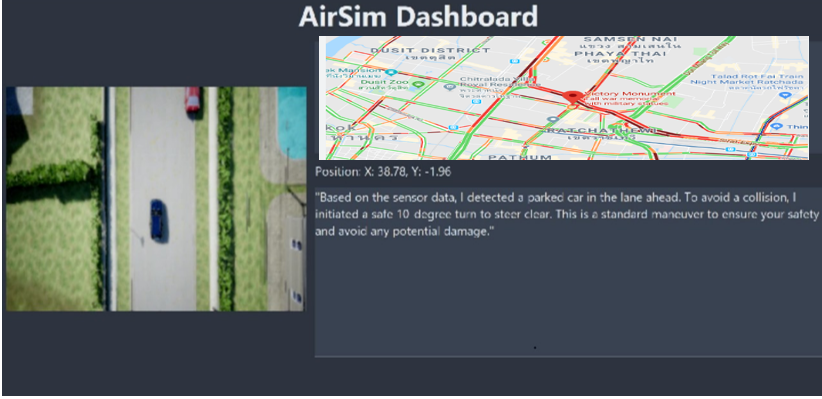}
    \caption{Example of HMI Visual Interface}
    \label{visual interface}
\end{figure}

BEV is a technology that provides an overhead perspective of the vehicle and its surroundings. It displays various objects, such as vehicles and pedestrians, from a comprehensive, top-down view. The primary purpose of BEV is to offer an integrated global view, enabling users to gain a thorough understanding of their environment and the dynamics of the situation. Additionally, BEV enhances spatial awareness, reduces blind spots, and provides improved visual aids, thereby increasing the user's sense of safety.

BEV has been shown to be a crucial tool for enhancing user situational awareness in autonomous driving systems. For instance, Cummings et al. (2010)\cite{cummings2010role} observed that bird's-eye views help users better understand object relationships within complex environments, while Palazzi et al. (2017) \cite{palazzi2017learning} demonstrated that bird's-eye views significantly reduce users' cognitive load.

Utilizing integrated map, including trip tracks and vehicle location, offers familiar geographic context, similar to Google Maps Navigation. Integrated maps provide accurate road and environmental information \cite{yang2024review}. This approach not only aids users in better identifying the vehicle's location but also enhances their trust in the current travel route.

The text display module provides detailed information about the current position and behavior of the vehicle and explains specific actions such as steering or stopping. By offering clear and concise textual explanations of the current surroundings and driving behavior, the interface helps users grasp the context and dynamics of the driving situation in relatable terms. Instant text information and explanations not only enable users to quickly access important information, but also help to increase their trust and comprehension of the driving trip \cite{tekkesinoglu2024advancing}.

\subsection{Voice Interaction with LLM}

Another crucial aspect of the HMI system is its voice interaction function, driven by a LLM. This feature utilizes the LLM to provide real-time language interpretation of vehicle behaviors and to respond to user queries. Autonomous driving system gives information about the driving environment and vehicle behaviors, including speed, acceleration, braking, direction changes, and so on. These explanations are also stored in our system’s memory for future reference.

\begin{figure}[ht]
    \centering
    \includegraphics[width=0.7\columnwidth]{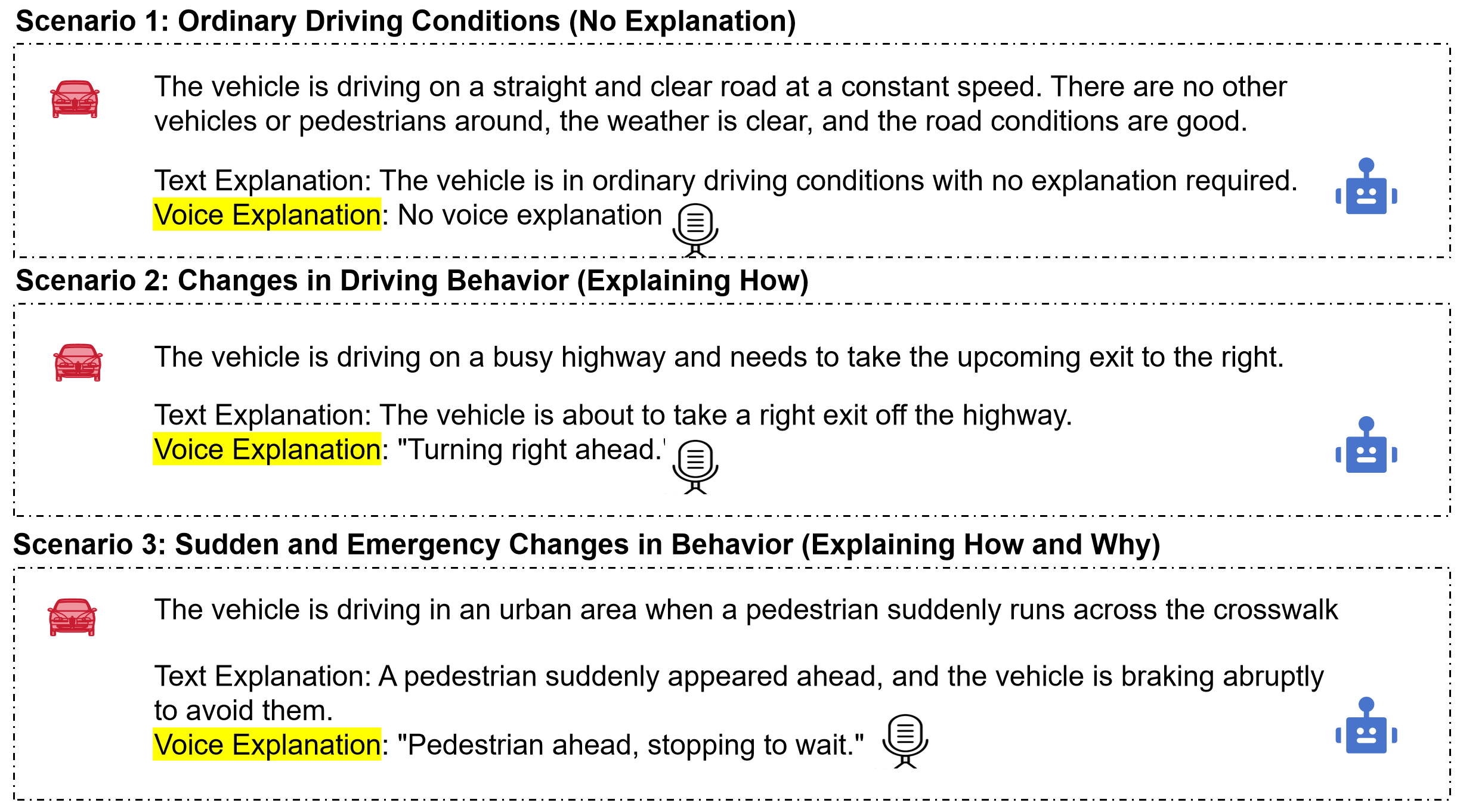}
    \caption{Examples of HMI Voice Interaction}
    \label{visual interface}
\end{figure}

We fine-tuned the speech output of a LLM through prompt engineering to prevent excessive noise during car travel. Research\cite{reis2009mobile} shows that an overload of speech content can lead to user irritation and anxiety, negatively impacting the overall experience. Consequently, we adjusted the LLM to ensure that the speech output remains concise and clear.

Our prompt engineering design is for meeting user needs for different explanations in various driving scenarios\cite{dong2023}. We have categorized the explanations into three scenarios:

\begin{enumerate}
    \item "No Explain": Applied under normal driving state when the vehicle’s behavior remains consistent.
    \item "Explain How": Used when the vehicle's driving behavior changes, such as during turns or lane changes. For example, the system will give a voice instruction like ``Turn right ahead.''
    \item "Explain How and Why": Employed during sudden or urgent changes, such as when avoiding a pedestrian. In this case, the system will alert with ``Pedestrian ahead, stop and wait.''
\end{enumerate}

This categorization ensures that the speech output is appropriate and necessary, preventing excessive information from disrupting the driving experience. It aims to enhance user acceptance and satisfaction with our voice interpreter.

This module also enables users to interact with our system using human language, and it provides responses in a conversational way. For example, When a user asks a question using voice commands, such as "Where are we now?" the system converts the spoken input into text, searches the LLM's memory for relevant information, formulates an appropriate response, and converts it back into voice to provide a verbal explanation to the user. Then our system will respond: “We are currently at the intersection of Main Street and Second Avenue, 2 kilometers from our destination.”

We prioritized the user experience during the design of the voice interaction. By providing clear, concise explanations and interaction in human language. our HMI system is more user-friendly, and effectively bridge the communication gap between the users and the autonomous driving technology.

\section{User Study}

We used an "experimental study" approach to investigate the effects of HMI systems on volunteers' trust in autonomous driving. This research paradigm is prevalent in human-computer interaction research and focuses mainly on assessing participants' performance and attitudes before and after product interaction. We started by recruiting 20 qualified volunteers. Volunteers were first asked to experience three different scenarios of autonomous driving without HMI interaction to collect their initial attitudes and evaluations of autonomous driving. After a day's interval, the same volunteers experienced three different scenarios of automatic driving with HMI interaction, collected their attitude and evaluation of automatic driving, and quantified the impact of HMI system by comparing the changes before and after.   

\subsection{Participants}
Participants were recruited using campus-wide emails, informational posters, and direct invitations during autonomous driving-related lectures and workshops. Interested individuals were required to sign an informed consent form outlining the study’s objectives, procedures, expected duration, and participant rights, including the right to withdraw at any time without penalty. We randomly selected 20 volunteers, balanced for gender with 12 males and 8 females, from 7 different countries and regions. Information distribution is depicted in \textbf{Table \ref{tab1}}. 
We chose people who were comfortable with autonomous driving to participate in the experiment. These volunteers represent a wide range of education levels, from doctoral to master's to undergraduate degrees. In order to ensure the diversity of recruitment groups, we used stratified sampling to select the same proportion of people from each education level to participate in this study, including 4 PhD students, 6 master students and 10 undergraduate students. After sampling, the sex ratio was adjusted to make it average, with 11 men and 9 women. In doing so, we ensured that the participants had a solid understanding of autonomous vehicles, were aware of the potential limitations of such systems, and frequently experienced riding in them in their daily lives. To address potential language barriers, the human-machine interface is programmed to operate in English. In addition, to ensure understanding and accessibility, each volunteer was provided with a translation of the interaction language prior to the experiment.

\begin{table}[h!]
\centering
\caption{Participant Demographics}
\label{tab1}
\renewcommand{\arraystretch}{0.8} 
\setlength{\tabcolsep}{4pt} 
\small 
\begin{tabularx}{\linewidth}{@{}c c c X c c@{}}
\toprule
Pair IDs & Gender & Age & Faculty & Degree & Location \\
\midrule
1 & M & 21 & Social Sciences & Bachelor & Canada \\
2 & F & 22 & Business & Master & CN \\
3 & F & 25 & Engineering & PhD & US \\
4 & F & 33 & Information Technology & PhD & US \\
5 & M & 18 & Science & Bachelor & Canada \\
6 & M & 29 & Social Sciences & PhD & HK \\
7 & F & 21 & Business & Bachelor & Malaysia \\
8 & M & 21 & Engineering & Bachelor & UK \\
9 & M & 30 & Information Technology & PhD & Australia \\
10 & F & 23 & Science & Bachelor & Malaysia \\
11 & F & 22 & Social Sciences & Master & HK \\
12 & F & 35 & Business & Master & Malaysia \\
13 & M & 19 & Engineering & Bachelor & UK \\
14 & F & 26 & Information Technology & Master & US \\
15 & F & 29 & Science & Master & Canada \\
16 & M & 20 & Social Sciences & Bachelor & Malaysia \\
17 & M & 19 & Business & Bachelor & Canada \\
18 & M & 19 & Engineering & Bachelor & UK \\
19 & M & 21 & Information Technology & Bachelor & Australia \\
20 & M & 31 & Science & Master & US \\
\bottomrule
\end{tabularx}
\end{table}

\subsection{Experimental design}
The simulation environment was rendered using Calar software, initially setting up three screens to replicate a passenger's perspective in a self-driving vehicle. We simulated three different scenarios: normal driving conditions with sunny weather and clear visibility; adverse weather conditions featuring rain and snow with reduced visibility; and an emergency scenario where a pedestrian unexpectedly appears on the road. All these simulations were conducted at a consistent speed of 30 kilometers per hour to reflect urban road conditions. In these simulations, the autonomous vehicle executed typical navigational maneuvers based on its algorithm.

In the subsequent comparative study, we integrated the HMI system to provide real-time explanations for passengers within CARLA-based simulated scenarios. After completing the experiments, passenger experience feedback was collected and evaluated.

\subsection{Procedure}
Prior to the experimental sessions, volunteers were invited to our simulation laboratory, where they received a briefing on the current capabilities and limitations of autonomous driving technology. This orientation included a video presentation that provided foundational knowledge of autonomous driving and a preliminary overview of the experimental procedures, aimed at minimizing potential biases due to varying levels of awareness among participants. After viewing the video, volunteers signed an approved human subjects consent form and proceeded to the simulated driving environment.

The experiment consisted of multiple scenarios, each lasting five minutes and spanning a distance of two kilometers, categorized into normal driving conditions, low visibility, and emergency situations. Between scenarios, to mitigate interaction effects and ensure the independence of each test condition, volunteers engaged in a 30-minute discussion on topics unrelated to autonomous driving.

Upon completing all scenarios, participants filled out a questionnaire evaluating their overall experience and their confidence in the autonomous technology. A follow-up session, conducted no sooner than 24 hours after the initial trial, involved a comparative experiment to assess the impact of the HMI system on user experience. This subsequent session strictly controlled for extraneous variables, maintaining consistency in procedural steps and experimental intervals. After each session, volunteers completed corresponding questionnaires, with results subsequently subjected to statistical analysis.

\subsection{Results}
We tested 20 volunteers and got 19 valid data points, one of whom dropped out of the test. We conducted qualitative and quantitative analysis of the data before and after. The results show that before using the HMI system, the number of people who said they did not trust autonomous driving (score three or less) was 47.4\%. In the tests conducted after using the HMI system, this percentage was reduced to 14.2\%, as shown in \textbf{Table \ref{tab2}}. This conclusion directly proves the effectiveness of the HMI system.

\begin{table}[ht]
\centering
\caption{Level of trust in Autonomous Vehicles (AV) before and after the implementation of the Multimodal Interpretation System (MI)}
\label{tab2}
\begin{tabular}{ccc}
\toprule
\textbf{Level} & \textbf{Before (\%)} & \textbf{After (\%)} \\
\midrule
1 & 10.5 & 0 \\
2 & 21.1 & 7.1 \\
3 & 15.8 & 7.1 \\
4 & 47.4 & 64.3 \\
5 & 5.3 & 21.4 \\
\bottomrule
\end{tabular}
\end{table}

Under Ordinary driving conditions, the implementation of the HMI significantly increased user trust in autonomous vehicles, with the proportion of participants expressing "trust" and "complete trust" rising by over 30\% to reach 85.7\% as shown in \textbf{Table \ref{tab3}}. However, the recognition of the Longitudinal Lane Marking and BEV systems varied across different driving conditions. Participants generally found the LLM explanations more trustworthy in extreme conditions, with trust in the LLM increasing by 12\%, 24\%, and 3\% under various challenging scenarios. Conversely, in low visibility conditions, initial trust in the BEV system decreased after the introduction of HMI, with a decline exceeding 20\%. These findings highlight the nuanced impacts of HMI on trust in different driving environments, emphasizing the importance of context-specific multimodal feedback to enhance user confidence in autonomous driving technologies.
\begin{table}[h!]
    \centering
    \caption{Performance of LLM and BEV Methods in Different Conditions}
    \label{tab3}
    \begin{tabular}{lccc}
        \toprule
        \textbf{Method} & \textbf{\shortstack{Ordinary \\ Environment}} & \textbf{\shortstack{Low \\ Visibility}} & \textbf{\shortstack{Emergency \\ Braking}} \\
        \midrule
        LLM (before) & 73.70\% & 68.40\% & 89.50\% \\
        LLM (after) & 85.70\% & 92.90\% & 92.90\% \\
        BEV (before) & 68.40\% & 73.70\% & 73.70\% \\
        BEV (after) & 71.40\% & 50.00\% & 71.40\% \\
        \bottomrule
    \end{tabular}
\end{table}
After actually using the HMI system, we found that the level of trust in the LLM and BEV changed, because in the previous test, the volunteers did not actually use the LLM and BEV for environmental interpretation. After the use of HMI, the LLM explanation was more trusted by the volunteers. The volunteers significantly increased their trust in the LLM interpretation in all settings, as shown in \textbf{Table \ref{tab4}}. 

\begin{table}[h!]
    \centering
    \caption{Surveys that Enhance User Trust}
    \label{tab4}
    \begin{tabular}{lccc}
        \toprule
        \textbf{Method} & \textbf{\shortstack{Ordinary \\ Environment}} & \textbf{\shortstack{Low} Visibility} & \textbf{\shortstack{Emergency \\ Braking}} \\
        \midrule
        Only LLM & 28.60\% & 21.40\% & 14.30\% \\
        Only BEV & 7.10\% & 7.10\% & 7.10\% \\
        Both & 64.30\% & 71.40\% & 78.60\% \\
        \bottomrule
    \end{tabular}
\end{table}
Different driving environments influenced the volunteers' response to using HMI, and as the driving environment became more complex, volunteers preferred to use a hybrid interpreter of BEV and LLM.

\subsection{Limitations}
Our study is constrained by its reliance on simulators.  While these simulators capably reproduce various weather conditions and autonomous driving behaviors and provide explanations, they cannot fully replicate the complexities of real-world driving environments. Research conducted in actual autonomous driving contexts would likely yield more accurate measurements of changes in trust levels.
Additionally, our participant sample presents demographic limitations, primarily comprising young to middle-aged individuals affiliated with colleges and universities.  This demographic skew may overlook significant variations in trust that could manifest among elderly populations or individuals with lower educational levels, who were not represented in our study.

\section{Conclusion}
In this paper, we designed a HMI system combining a visual interface integrated with BEV, Map and Text Display as well as a prompt engineering fine-tuned LLM as voice interaction. Furthermore, we tested HMI for user study in an experimental autonomous driving environment, and the results demonstrated its effectiveness in enhancing user trust. These findings indicate that our HMI can significantly enhance user confidence in autonomous vehicles and adapt well to diverse driving conditions, making it a valuable tool for future autonomous driving systems.

{\small
\bibliographystyle{ieee_fullname}
\bibliography{paper}
}

\end{document}